# OF CATS AND QUANTA: PARADOXES OF KNOWING AND KNOWABILITY OF REALITY

Research Paper

Gennady Shkliarevsky
Department of History
Bard College
Annandale-on-Hudson, NY 12572
USA
Tel.  845-758-7237 (office)
845-876-3091 (home)
shkliare@bard.edu

Introduction

Paradoxes have always been and remain a source of fascination and anxiety.  On one hand, their enigmatic nature challenges our creativity and inspires our imagination.  On the other hand, paradoxes appear to be a compelling evidence of some fundamental structural limitations of our capacity to know.[1]

The interest in paradoxes goes all the way back to the early periods of human civilization.  All major philosophers of Ancient Greece were keenly interested in paradoxes.  The so-called "insolublia" were extremely popular during the Scholastic age.  Modern age thinkers held an equal admiration for paradoxes.  Paradoxes also continue to attract much attention in our time.

The source of this enduring interest is the fact that in a deceptively simple way paradoxes raise one question that is crucially important to human civilization:  Are there any inherent limitations to human thought?  Is there anything that we, humans, absolutely cannot know as a matter of principle, rather than as a result of a flawed approach?  Some of the best minds in the history of humanity have tried to solve the mystery of paradox.

Yet paradox has successfully resisted all attempts at understanding its nature; its secrets remain as impenetrable today as they have been throughout centuries.

Definitions of paradox are numerous and diverse. They range from simple to complex, from descriptive to analytical, from intuitive to formal. In his recent book on the history of paradox, Ray Sorensen, who traces paradoxes to the riddles of Greek folklore, describes them as "questions . . . that suspend us between too many good answers."[2] R. M. Sainsbury defines paradox as "an apparently unacceptable conclusion derived by apparently acceptable reasoning from apparently acceptable premises."[3] In her exploration of paradoxes, Marianne Lewis writes: "'Paradox' denotes contradictory yet interrelated elements—elements that seem logical in isolation but absurd and irrational when appearing simultaneously."[4]

There seems to be one distinct feature to which many definitions of paradox point: paradoxes are about truths that are true but their truth is impossible to prove. One of the most famous of all paradoxes, the Liar Paradox, is a good illustration of this point.[5] The Liar tells the truth only if he lies, and lies only if he tells the truth. The Liar seems to be doing the impossible from the point of common logic: he tells truth and lie at the same time.

Many thinkers have made attempts at understanding the source of paradoxes. There are two principal positions on this point. Both of them reflect the traditional dualistic approach. According to one of these positions, which may be called realist, paradoxes are rooted in the paradoxical nature of reality which is grounded in tensions and contradictions. The other position, which may be called constructivist, argues that paradoxes are products of human thought. Heraclites, for example, argued that paradoxes were part of reality; they were due to the actual existence of many truths. Parmenides, on the other hand, attributed paradoxes to human frailties and inconsistencies which resulted from the reliance on imperfect and deceptive senses. Aristotle also regarded paradoxes as errors in human thinking, either involuntary or intentional. In his view, paradoxes were either fallacies or rhetorical devices that could be used in disputations in order to confuse the opponent. Rationalists saw paradoxes as a product of human failure to have adequate <u>a priori</u> insights. Empiricists insisted that they were due to riddles of nature itself that our mind simply could not resolve. Humans, they argued, must have wisdom and maturity to recognize those riddles which the human mind could understand and those about which it would remain forever ignorant. For Existentialists, the very tensions and contradictions of human existence generated paradoxes.

The dichotomy between realism and constructivism is also present in the current discussions of paradox. In his study of paradoxes, Roy Sorensen has argued in favor of fundamental cognitive flaws as the source of paradoxes.[6] By contrast, R. M. Sainsbury has been critical of the emphasis on overarching cognitive patterns in explaining paradoxes and has been forcefully arguing in favor of realism and the actual existence of many truths.[7]

Philosophers have not been the only ones who have been interested in paradoxes. Many disciplines display a keen interest in this subject. Psychology, for example, pays much attention to cognitive aspects of paradoxes. Organizational Studies are also vitally interested in understanding paradoxes which are viewed as "contradictions embedded within a statement, human emotions, or organizational practices." Rather than regard paradoxes in a negative light—as a problem that needs to be resolved—many disciplines view paradoxes in a positive way as crucial factors in the construction of knowledge. Some actually embrace paradoxes and focus on their production and utilization. Ford and Backoff, for example, see that paradoxes can be used for advancing knowledge. In their view, paradoxes can be constructed by bringing "oppositional tendencies . . . into recognizable proximity through reflection of interaction."[8] Sainsbury also associates paradoxes with crises and revolutionary advances in thought.[9]

Kurt Gödel, the celebrated Austrian logician and mathematician, has made probably the most important and far-reaching contribution toward our understanding of paradoxes. In 1931 Gödel published a ground-breaking paper that proved to be in many ways revolutionary. Gödel wrote this paper to address what was then one of the major problems in mathematics and logic—the problem of consistency. As formulated by the German mathematician David Hilbert the question that Gödel addressed was briefly this: Is it possible to construct an absolute proof of consistency of an axiomatic system? That is, is it possible to construct a proof that all true statements in a given system are consistent with its axioms?

In his article Gödel proved in a very ingenious and absolutely incontrovertible way that such proof was in principle impossible. He showed that any deductive system could have sentences which, although true, were indemonstrable; in other words, their truth could not be proved within the given set of axioms and rules of inference and hence, from the perspective of this system, they were not true. In order to demonstrate their truth, one would have to resort to meta-mathematical procedures and construct a new and broader axiomatic structure which was sufficiently powerful to construct such proof. However, even a new structure would still allow a possibility of constructing new sentences which, although true, could not be proven within this structure. Gödel's proof showed the inherent limitations of the deductive method, which was and continues to dominate the sciences as the principal method for demonstrating truth. Gödel's proof suggested a paradoxical conclusion that was encouraging and frustrating at the same time: although the power of human mind to generate knowledge seemed to be infinite, it could not devise a system for controlling this power.[10]

Among many other things, Gödel's contribution has interesting implications for our understanding of the nature of paradox. For example, Gödel has shown that within any cognitive structure[11] it is possible to construct knowledge that transcends this structure and cannot be controlled by it. In order to control this knowledge one has to construct a new and more powerful structure. In other words, what Gödel's contribution suggests is that paradoxes emerge when two cognitive structures—one more powerful than the other—intersect and the less powerful cognitive structure is perturbed by the emerging more powerful one which is capable of formally proving the controversial truths indemonstrable within the old structure. Therefore, an understanding of paradoxes and their function is closely related to the study of cognitive structures and their evolution.

It is beyond the scope of this paper to deal with the evolution of cognitive structures. I would direct the reader to the works of Piaget and my own discussion of this topic.[12] Rather, what I would like to do in this paper is to concentrate on a clash between cognitive structures in a real paradox.

The Paradoxical Universe of Modern Physics

Paradoxes are abundant in contemporary physics. One could start with the fact that it provides two very different and largely incompatible descriptions of physical reality—one by general relativity (GR) and another by quantum theory (QT). The principal approaches and physical laws that operate in one description are largely inapplicable in another. QT practically ignores the principle of invariability of frames of reference that is central to GR. Phenomena, such as non-locality, entanglement, superposition, complementarity, randomness, which are predicted and verified by QT, are unobservable in GR.

The universe described by QT appears to make absolutely no sense when viewed outside its formalism. For example, how can one make sense of non-locality which involves speeds faster than the speed of light, an absolute constant in GR? Or, what should one make of superposition, according to which a quantum system can be in two different states at the same time. The contradictions with our familiar sense of how physical reality operates are so great that even many who are intimately familiar with QT find its puzzles hard to comprehend. Richard Feynman, who received a Nobel Prize for his achievements in QT, cautioned:

> Do not keep saying to yourself, if you can possible avoid it, "But how can it be like that?" because you will get 'down the drain,' into a blind alley from which nobody has yet escaped. Nobody knows how it can be like that.[13]

Of all the paradoxes associated with QT, the paradox of Schrödinger's cat is one of the most famous. In 1935 the famous Austrian physicist Arnold Schrödinger formulated the following <u>Gedankenexperimente</u>: Imagine a cat sitting inside a hermetically sealed box which is connected to a container with a poisonous gas. The valve in this connection is controlled by a quantum system: if the system is in one state, the valve remains closed and the cat stays alive; if the system is in another allowable state, the valve opens and the cat dies. The problem is, however, that according to QT, a quantum system can be in two different superimposed states at the same time. Therefore, when the system is in such state, the cat should also be alive and dead at the same time. It is all right to accept the mathematical formalism that allows quantum system to be in two different states at the same time, but one finds it very difficult to accept a possibility of a cat being in two such different states as alive and dead at the same time.

Numerous attempts to provide an alternative to what is called the standard or Copenhagen interpretation of QT, which was formulated by Niels Bohr, Werner Heisenberg, Max Born, and Wolfgang Pauli, have addressed this <u>Gedankenexperimente</u>, including the many-worlds interpretation (Hugh Everette and John Wheeler), the hidden-variable interpretation (Frederik Belinfante and Max Jammer), the "disturbance model" (Nick Herbert), the alternative collapse model, or rather models (John von Neumann, Fritz London, Ernest Bauer, and Eugene Wigner), advanced-action interpretations, and others. It is beyond the scope of this paper to go into a detailed analysis of these interpretations.[14] It is worth mentioning, however, that all these attempts ended with interpretational problems of their own.

In order to resolve a problem, one should try to understand its source. There are two very distinct entities involved in the paradox of Schrödinger's cat: the cat and the decaying particle. The problem is that they belong to two very different domains recognized by contemporary physics, and the phenomena of the quantum micro domain appear to be absurd from the point of view of the macro domain. It is the clash between these two domains which creates the paradox of Schrodinger's cat. Thus the source of this paradox is the dualistic distinction between the macro and micro domain which is characteristic for the contemporary descriptions of physical reality, and the best illustration of this distinction is the conflict between theory of relativity, which is the main theory for describing the domain of macro phenomena on the cosmic scale, and quantum theory, which provides the principal description of the micro phenomena of the quantum universe. The two theories do not use the same formalism, they recognize different laws of nature, and in general largely do not talk to each other.

So, the original question of the source of the paradox of Schrodinger's cat is transformed into another question: Why is there this dualistic distinction between macro and micro domains in contemporary physics? What is the source of this dualism?

Dualism is certainly not unique to contemporary physics. Many other spheres of human knowledge reflect the dualistic perspective. Dualism in the European intellectual tradition has a long history. Arguably, the oldest and the most enduring dualism in this tradition is that between the subject and the object which survives to this day. Dualism is also not unique to Europe but in fact is an integral part of many other intellectual traditions.

The dualistic perspective reaches far beyond the mere philosophical problems of subject-object relations and affects the very view of knowledge and knowledge acquisition, differentiating it into two distinct and largely mutually exclusive perspectives—rational and irrational. Although most epistemologies tried at least to recognize the validity of both approaches, dualism has successfully resisted any productive synthesis. Most attempts at synthesis generally place emphasis on one approach at the expense of another.  One can see dualism at work, for example, in the European tradition in the distinction made between reason and faith.

It is interesting that although the European intellectual tradition has experienced many changes in the course of its evolution, dualism in it has remained largely intact. Secular knowledge that emerged during the Early Modern period represented in many ways a profound shift in its conception of reality by comparison with the Middle Ages. One of the results of this shift was the emergence of modern science with its analytical reductionist approach and atomism, as opposed to the holistic approach with an emphasis on faith which was predominant during the Middle Ages.  However, its conception of knowledge production remained essentially unchanged.  Just like in the Middle Ages, knowledge was not regarded as a product of construction, but rather as a more or less passive reflection of reality.  The process of the construction of knowledge was not recognized and, as a result, the focus of this perspective remained largely, as in Antiquity and the Middle Ages, on the object which was deemed to be completely separate from the largely passive subject.

The focus on the object certainly did not eliminate the real activity of the subject in constructing knowledge; it merely rendered this activity invisible and hence uncontrollable.  Rather then offer a critical control over the activities of the subject, this approach involved an unconscious and uncritical projection of the subject on the object. It is due to this unconscious projection that classical physics accepted the frame of the preferred knower—for example, a physicist like Newton—as the absolute frame, or absolute space.

Since the space constructed by the preferred knower was regarded as absolute, all observed and even unobserved events had to be located in this space. Hence, simultaneity of events could be established which led to the acceptance of the notion of absolute time. Also, as a result of the uncritical projection of the agency of the knower on objects and their relations, the necessity of the agent's action in constructing knowledge was projected onto relations among objects as necessary and deterministic, which gave rise to the largely anthropomorphic notion of laws of nature. Time inversion was another consequence of the unconscious and uncritical modeling of spatio-temporal relations on the activity of the agent. Finally, since the classical perspective knew only one preferred frame and since it did not know any speeds faster than the speed of light, it recognized the principle of locality which asserted that causal relations could not exceed the speed of light.

The beginning of the 20$^{th}$ century witnessed a radical transformation of modern physics. The two most important innovations were the theory of relativity formulated largely by Albert Einstein and quantum theory of the Copenhageners. The recognition of the agency of the knower was central to both innovations. The introduction of the point-of-view invariance for the frame of reference was seminal for TR. In Einstein's view, space should look invariant regardless of the frame chosen by the knower. Einstein's dictum was that no frame should be given preference. This central tenet contained a powerful recognition that all frames are constructed and therefore all are equal. The only non-relativistic component in Einstein's picture of the universe was light. The speed of light had to be the same in all frames, and therefore constant. If it were not, then some frames had to be different from others.

One of the logical consequences that followed from the acceptance of the speed of light constant was Einstein's insistence on the classical principle of locality in his view of causality. No causal relations in the universe should exceed the speed of light. While Einstein recognized the agency of the knower in choosing/constructing frames of reference, his theory in general did not consistently incorporate the recognition of the process of construction. As a result, despite its significant differences from classical mechanics, theory of relativity retained dualism. This dualism manifested itself, for example, in Einstein's insistence on the classical separation between subject and object. Just like Newton, Einstein believed that it was possible to observe reality "without in any way disturbing the system" under observation. Thus, despite his recognition of the role of agency in constructing frames, Einstein still largely associated knowledge with passive observing.

QT also marked a radical departure, and probably even more radical than that of TR, from classical mechanics. It was much more consistent than the latter was in recognizing the agency of the knower. It no longer viewed the knower as a passive observer but rather as an active agent whose interaction with the object radically changed it. According to QT, for example, the knower radically affected, one could even say produced, the outcome of experiments (for example, measurements performed on a particle).

However, while QT recognized the agency of the knower, their description of reality still included a classical component. The inclusion of the classical component related primarily to the conditions of the experiment. It was with regard to the conditions of the experiment that the Copenhageners insisted on a classical description. As Bohr stressed, ". . . it is decisive to recognize that, <u>however, far the phenomena transcend the scope of classical physical explanation, the account of all evidence must be expressed in classical terms</u>."[15]

There are several interpretations as to why the Copenhageners included the classical component in their description of reality, including one by the Copenhageners themselves. Echoing the Copenhageners, Landsman argues in favor of pragmatic considerations, while Henry Stapp's explanation stresses mere inertia.[16] As I have argued elsewhere,[17] the reasons for the inclusion appear to have more to do with the fact that although QT recognizes the relationship between the subject and the object, their epistemology does not fully integrate this recognition. As a result, the process of the actual interaction between the subject and the object remains in QT just as mysterious as it is in classical mechanics; the only real difference between the two is that QT focuses on the subject while the focus in classical mechanics is on the object.

In the absence of a clear understanding of the interaction between the agent and the object, the focus on the knower rendered the object vague to the point of being a mere abstraction, a kind of Kantian "thing-in-itself." Schrödinger aptly described quantum mechanics as "a formal theory of frightening, indeed repulsive, abstractness and lack of visualizability."[18] As a result of their recognition of the agency of the knower, the Copenhageners faced a very serious problem: their description of reality could be regarded as unacceptably subjective. They needed to bring a degree of objectivity into their description and the only part of their description where they could do it was in their description of the aspect related to the knower. Since they knew only one way to bring in objectivity—the classical way that posits separation between the observer and the

observed—they insisted on providing a classical description of the conditions of the experiment, that is, presuming that these conditions exist separately from the knower and that the knower "in no way affects" the description of these conditions.

Thus classical epistemology made its way into the new quantum description of reality. As a result, the knower was displaced into a transcendent, critically uncontrolled, and preferred frame. The consequence of this displacement was, just as in classical mechanics, the recognition of the frame constructed by the knower as the absolute frame for quantum events.[19] Thus quantum theory abandoned the point-of-view invariance with regard to space incorporated by Albert Einstein into his theory of relativity. The introduction of the absolute space permitted the establishment of simultaneity between quantum events which logically led to the recognition of absolute time for quantum events. As Dirac has observed, "It [QT] is against the spirit of relativity, but it is the best we can do. . . . We cannot be content with such theory."[20]

The quantum universe with its absolute space-time has become a very odd place indeed: it allows speeds faster than the speed of light, permits systems to be in two different states at the same time (superposition), reverses the arrow of time and causality or even allows events that do not seem to have any cause at all, and last but not least, presents numerous examples of non-local behavior, or what Einstein sarcastically called "a spooky action at a distance." It is a universe of uncertain, random, indeterminate, and entangled events. From the point of view of the physical universe which we experience in our daily life, quantum events appear to be paradoxical and even absurd. As Richard Feynman once remarked, "I think it is safe to say that no one understands quantum mechanics."[21] The paradoxes accepted in quantum mechanics vividly exemplify a radical division of physical reality into two distinct and largely mutually exclusive domains and a resurgence of traditional dualism in contemporary physics.

Although over the years QT has led to many remarkable discoveries, it also has been and continues to be plagued by numerous problems. Many in the field still wonder what kind of universe lurks behind its mathematical formalism and try to lift what they call "the fog from the north," as they refer to the Copenhagen interpretation. Adrian Kent articulates a very common concern when he writes:

> We know [from CI] that microscopic systems behave in a qualitatively different way, there is intrinsic randomness in the way they interact with the devices we use to probe them. Much more impressively, for any given experiment we carry out on microscopic systems, we know how to list the possible outcomes and calculate the probabilities of each, at least to a very good approximation. What we do not fully understand is why those calculations work: we have, for example, no firmly established picture of what (if anything) is going on when we are not looking.[22]

According to John Cramer, the author of the transaction interpretation of QT, the greatest weakness of QT

> . . . is not that it asserts an intrinsic randomness but that it supplies no insight into the nature or origin of this randomness. If "God plays dice," as Einstein has declined to believe, one would at least like a glimpse of the gaming apparatus that is in use.[23]

One clearly senses frustration in the words of the Russian physicist Lev Chebotaryov who laments the fact that seventy years after the advent of QT, "there is still no clear idea as to what its mathematics is actually telling us."[24]

It is beyond the scope of this paper to discuss in detail numerous attempts to reinterpret and bring more clarity to QT. A cursory discussion of even the most important of these reinterpretations may easily be a topic for a book-length project. It is worth, however, to mention a few alternatives to the Copenhagen interpretation, such as the hidden variables interpretation, the many worlds interpretation (Hugh Everette),[25] the stochastic interpretation (Louis de Broglie, David Bohm, and Jean-Pierre Vigier),[26] the delayed choice interpretation (John Wheeler),[27] and interpretations involving information theory. However, despite these numerous efforts and many remarkable discoveries, the picture of nature as described by QT remains quite "foggy."

The continued split between TR and QT also presents problems for many physicists. How can there be two physical realities? How do they relate to each other? And where is the line that separates them? In his well-known article appropriately titled "Do we really understand quantum mechanics?" Franck Laloë writes:

> Logically, we are faced with a problem that did not exist before, when nobody thought that measurements should be treated as special process in physics. We learn from Bohr that we should not try to transpose our experience of the everyday world to microscopic systems; this is fine, but where exactly is the limit between the two worlds?[28]

Many physicists see an urgent need for a synthesis of the two most important theoretical perspectives in modern physics. The following reflection offered by John Baez summarizes the attitude of many in the field:

> General relativity and quantum field theory are based on some profound insights about the nature of reality. These insights are crystallized in the form of mathematics, but there is a limit to how much progress we can make by just playing around with this mathematics. We need to go back to the insights behind general relativity and quantum field theory, learn to hold them together in our minds, and dare to imagine a world more strange, more beautiful, but ultimately more reasonable than our current theories of it.[29]

Adrian Kent echoes the same sentiment when he writes: ". . . almost everyone suspects that a grander and more elegant unified theory . . . await us."[30]

Transactional Interpretation and the Legacy of Atomism

Although over the years there have been many attempts to reconcile QT and TR, the two theoretical perspectives remain deeply divided. It is worth taking a closer look at least at one such very interesting attempt as it may help to understand better reasons for these failures. In 1986 John G. Cramer, a physicist from the University of Washington, published the article entitled "The transactional interpretation of quantum mechanics" in the journal Review of Modern Physics.[31] In this article Cramer advances a new interpretation of quantum formalism. The new interpretation pursues two principal objectives. First of all, Cramer wants to reconcile QT and TR by introducing into the quantum domain one of the main features of TR—the principle of invariability of space frame, that is, Einstein's condition <u>sine qua non</u> that the universe should look the same no matter what is the chosen frame of reference. In Cramer's own words:

> The transactional interpretation of quantum mechanics . . . is based on solutions of relativistically invariant differential field equations, is fully consistent with special relativity, and seems to accommodate these additional features of a relativistic quantum theory in a very natural way. We are therefore confident that the interpretation presented here, perhaps with minor embellishments, is appropriate for the interpretation of a fully relativistic theory of quantum mechanics.[32]

Secondly, Cramer wants to provide an interpretation that would go beyond the Copenhagen interpretation and offer a "conceptual model which provides the user with a way of clearly visualizing complicated quantum processes and of quickly analyzing seemingly 'paradoxical situations." He makes it absolutely clear that his interpretation differs from the standard one in a "way of thinking rather than in a way of calculating." He sees particular utility of his interpretation as a pedagogical device in helping students to visualize complex quantum processes.[33]

In contrast to the standard interpretation which is predicated on the absolute space frame, Cramer recognizes that the emitter and the absorber have different frames. The recognition of two different frames—that is, the recognition of the non-local nature of the interaction between emitter and absorber—poses a problem in explaining the enforced correlation in polarizations in the two frames. According to Cramer, this correlation is achieved via a standing wave formed by the superposition of a retarded, or in his terminology offer wave (OW) from emitter to absorber, and the advanced, or confirmation wave (CW) from absorber to emitter). Cramer provides a detailed description of the causal sequence in this superposition of waves. First, emitter produces an offer wave, which travels to absorber. When it arrives, it causes absorber to produce a confirmation wave. This wave travels back to emitter where it is evaluated. This cycle is repeated until the "exchange of energy and other conserved quantities satisfies the quantum boundary conditions of the system," at which point the transaction is completed.[34]

There is one problem with this description: all these sequential exchanges of offering and confirmation waves take place before an observer can see the complete transaction which is interpreted "as the passage of a single retarded (i.e., positive-energy) photon traveling at the speed of light from emitter to absorber."[35] Therefore, they should occur at speeds that exceed the speed of light, which contradicts theory of relativity. So Cramer faces a dilemma: either he should abandon quantum non-locality or accept phenomena that are incompatible with the theory of relativity. Cramer is certainly not unaware of this tension. He writes:

> It is perceived by some that non-locality must be in direct conflict with special relativity because it could be used at least at the level of Gendankenexperimente" for "true" determinations of relativistic simultaneity and must be in conflict with causality because it offers a possibility of backward-in-time signaling.[36]

Cramer's solution to this problem is atemporality. He argues that transaction exchanges take place outside of time:

> Since the transaction is atemporal, forming along the entire interval separating emission locus from absorption locus "at once," it makes no difference to the outcome or the transactional description if separated experiments occur

"simultaneously" or in any time sequence. Both measurements participate equally and symmetrically in the formation of the transaction.[37]

Thus Cramer salvages both quantum non-locality and relativistic invariance. His interpretation "is explicitly non-local but it is also relativistically invariant and fully causal."[38]

This explanation, however, raises one problem: if transaction exchanges take place outside of time, they are certainly not outside of space. Because they are outside of time, they appear to the observer as instantaneous. This fact allows the observer to establish the fact of simultaneity between two different space frames—one of emitter and the other of absorber—which contradicts the theory of relativity explicitly precluding such possibility. An attempt by the Russian physicist Pavel Kurakin and his colleagues George Malinetskii and Howard Bloom to build on Cramer's interpretation has hardly been more successful. They proposed a "conversational," or "dialogue model of quantum transitions."[39] Here is the crux of their interpretation in their own succinct summary:

> We propose that the source of a particle and all of that particle's possible detectors "talk" before the particle is finally observed by just one detector. These talks do not take place in physical time. They occur in what we call "hidden time." Talks are spatially organized in such a way that the model reproduces standard quantum probability amplitudes.[40]

However, their proposed "hidden time" hardly resolves the contradiction between non-locality and relativistic invariability.

It is worth noting that Cramer's interpretation, despite its obvious innovations, remains decidedly traditionalist in its atomistic approach, that is, he sees quantum interactions as essentially interactions between two individual quantum systems. The reductionist atomistic approach and its rational analytical methodology re-emerged during the Early Modern period largely in opposition to the holistic approach and reliance on faith prevalent during the Middle Ages. Although this approach is in many respects very different from the medieval holistic perspective, there is one aspect that the two approaches share: they both set the part and the whole in a binary opposition to each other, except that the medieval perspective emphasizes the whole over the part, and the modern scientific perspective emphasizes the part over the whole.

The failure to achieve the integration of contemporary physics within the framework of the atomistic reductionist approach suggests the necessity of a philosophical rethinking of the applicability of this approach. John Small voices the opinion of the growing number of physicists when he writes:

> We need to go back to the insights behind general relativity and quantum field theory, learn to hold them together in our minds, and dare to imagine a world more strange, more beautiful, but ultimately more reasonable than our current theories of it. For this daunting task, philosophical reflection is bound to be of help.[41]

Self-Organization: a New Paradigm

As the above discussion of paradoxes in contemporary physics suggests, these paradoxes are in a large degree due to the failure of QT and TR to integrate fully the subject into their description or reality. As the above also indicates, and as I have argued elsewhere relying largely on the work of Swiss psychologist Jean Piaget,[42] the subject and the object are intimately related to each other by the process of construction. Piaget was one of the first thinkers who focused on this process. According to Piaget's many remarkable studies, and most notably his early empirical work on the origin of intelligence in children, [43] the subject and the object are not givens; they are products of our construction of knowledge about reality. It is the process of construction that is the source of both. And it is this process that integrally connects both the subject and the object which constitute the two poles of this continuum. Therefore, when we disregard the process of construction and focus our attention on either or both of the two poles—the products of construction—they naturally appear to us as disconnected and radically distinct.

The focus on the subject and the object has been and continues to be dominant in our view of reality at the expense of the process of construction. As a result, in this view the subject and the object remain radically separated from and largely opposed to each other. This focus was and continues to be the source of enduring dualism in our thinking. Therefore, the resolution of the problem of paradoxes in contemporary physics should involve, as John Small suggests, a philosophical re-thinking of its epistemological approach. The shifting of the focus to the process of construction may be instrumental to such re-thinking. One of the most important and productive among modern theoretical perspectives which focus on the process of construction is theory of self-organization.

Although theory of self-organization is not yet widely acceptable in QT, it is not entirely alien to it either. In fact, QT has always regarded quantum objects as complex dissipative systems. Both Bohr and Heisenberg referred to quantum systems when describing interactions in quantum mechanics. A description of quantum non-locality by the Indian physicist Ashok Sengupta is very reminiscent of the language used in theory of self-organization:

> Quantum non-locality is a natural consequence of quantum entanglement that assigns multipartite systems with definite properties at the expense of the individual constituents thereby rendering it impossible to reconstruct the state of a composite from knowledge of its parts.[44]

The Russian physicist Vladimir Manasson asks in the title of his paper: "Are Particles Self-Organized Systems?"[45] Chinese physicist Xiao Gang Wen thinks in a similar vein when he writes:

> Why don't we regard photons and fermions as emergent quasi-particles like phonons? It seems there does not exist any order that gives rise to massless photons and nearly-massless fermions. This may be the reason why re regard photons and fermions as elementary particles.[46]

Other physicists make similar connections between quantum processes and theory of self-organization.[47] Pavel Kurakin and George Malinetskii, for example, even go so far as to evoke analogies between paradoxes of quantum mechanics and the processes of self-organization among bees.[48]

The term "self-organization" refers broadly to processes which lead to the emergence of entities with properties that cannot be traced to individual parts involved in their emergence. Iain Couzin cites the following definition of self-organization provided by S. Camazine et al. In this definition, self-organization is

> . . . a process in which pattern at the global level of a system emerges solely from numerous interactions among the lower-level components of a system. Moreover, the rules specifying interactions among the system's components are executed using only local information, without reference to the global pattern.[49]

As the computer scientist James Odell succinctly puts it, an emergent complex system "is more than just the sum of its parts."[50] Theory of emergence and the concept of non-linearity are also commonly used in describing self-organized complex systems[51]—a view that goes back to Michael Polany's theory of boundary conditions that form constraining regulating processes.[52] Finally, self-organized system are also often characterized as dissipative—a characteristic that stresses their capacity to increase entropy in their environment while sustaining and developing their own organization.

Theory of self-organization is inherently interdisciplinary. It is intimately related to several theoretical perspectives, such as theory of catastrophe, chaos theory, theory of complex dissipative structures, theory of autopoiesis, theory of complexity, emergence theory, and information theory.[53] All these theoretical perspectives focus on a broad range of phenomena that occur both in the macro and the micro domain and cover many disciplines which study different levels of the organization of reality: from physics and chemistry to meteorology and weather science, to computer science, biology, psychology, economics, sociology, and linguistics. The theory of self-organization underlies, for example, studies of such diverse phenomena as collective behavior of animals,[54] insect behavior and swarm intelligence,[55] ball lightning,[56] weather patterns,[57] behavior patterns of fish schools,[58] and even linguistic processes.[59]

The emergence of new properties in self-organized systems attracts much attention. Many researchers see self-organization as the source of these properties whereby the interaction of individual parts on the local level leads to the emergence of global patterns constituting new properties. Jean Piaget shows, for example, how the mechanism of self-organization operates in mental processes. His study of the emergence of conscious intelligence in children demonstrates how conservation of individual reflex functions, such as hearing or seeing, leads to their combination and the emergence of a more powerful operation capable of creating permanent mental images, which is the first step in the rise of consciousness. Among other things, Piaget's study also shows that the part and the whole, which are commonly regarded as standing opposed to each other, are simply aspects of the same process of construction and do not stand in radical opposition to each other. [60] Individual functions interact in creating a whole that is not reducible to these functions. This whole, or the new function, is much more powerful than each individual function involved in the combination. This combination is capable of a number of operations that exceeds the number of operations that can be performed by the mere sum of individual operations involved in the combination. The transition from individual functions to a whole seamlessly combines both continuity and discontinuity. It is only when we remove from our field of vision the process of construction that part and whole, continuity and discontinuity, and other binaries begin to appear as opposed to each other.

    It is interesting to point out that in self-organized systems locality and non-locality are also not opposed to each other but reflect an integrated and cooperative aspect of the system's behavior. According to Eric Bonabeau, for example, self-organization involves

> . . . a set of dynamical mechanisms whereby structure appears at the global level of a system from interactions among its lower-level components; the rules specifying the interaction among the system's constituent units are executed on the basis of purely local information, without reference to the global pattern, which is an emergent property of the system rather than a property imposed upon the system by an external ordering influence.[61]

In his study of fish schools, Dmitrii Radakov also points to the connection between the local and the non-local in self-organized systems. He emphasizes that the behavior of such systems "need not be explained as a phenomenon coordinated by a leader, or by global information, but by the rapid propagation of local information about the motion of near neighbors."[62] He explains that the parts that constitute a self-organized system react to the behavior and position of their immediate neighbors, not to some "global information" or "leader," so information is passed locally in accordance with physical laws; however, due to self-organization, the effects of this locally passed information on the global behavior of the system appears to disobey physical laws.[63] The same pattern of local interaction with global effects is involved in the phenomenon which is called stigmergy whereby local information passed via pheromone affects the global behavior of an ant colony.[64] In linguistic studies, for example, Vito Pirelli and his collaborators show how the processing of verbal inputs through local interactions between parallel processing neurons leads to the emergence of global (non-local) ordering constraints.[65]

The peculiar connection between the local and the non-local in self-organized systems may explain, for example, the phenomena observed in self-organized quantum systems that appear to violate what Einstein considered to be one of the fundamental constants in nature—the speed of light—or the causal relationship whereby effects precede the cause or may even appear as having no cause at all, that is being totally random. Indeed, if local and non-local are intimately connected in self-organized systems and the effects of local interactions that occur in accordance with the laws of nature and its constants have global effects, then the effects of the local interaction on the global level may appear in violation of these laws and constants. In other words, events that happen on the local level at the speed of light and in accordance with natural laws are extended to the global level. In this case, their local effects are magnified due to self-organization and, naturally, would appear to violate these laws on the global level if self-organization is not taken into consideration.

In light of the peculiar effects of self-organization, one can understand the reason for the failure of the reductionist atomistic approach to resolve the paradoxes in quantum mechanics. Sengupta, for example, explains that

> A complex system behaves in an organized collective manner with properties that cannot be identified with any of the individual parts but arise from the structure as a whole: these systems cannot dismantle into their components without destroying themselves.[66]

Obviously, when the reductionist approach views the behavior of the system from the perspective of interaction between two individual and non-local particles, which in fact, do not interact directly but only as components the whole system, apparent paradoxes and inconsistencies will be an inevitable consequence.

The interdisciplinary nature of theory of self-organization suggests a certain isomorphism among different levels of organization of reality. Josephson, for example, notes deep parallels that appear to exist between patterns found at the physical level and patterns found in mental processes. Reflecting on the connection between two levels of organization, he writes:

> The details of quantum physics and biology are very different. But we argued that they might nevertheless be derivative of some common underlying subtler background process, in the same way that waves and particles emerge from a common subtler domain, that of quantum mechanics and in some cases share certain features such as propagation along a trajectory.[67]

Vladimir Manasson also points to the interdisciplinary nature of self-organization research when he writes that in order to "understand the overall SOS behavior, it is often sufficient to use qualitative analysis and study proto-type system that belongs to the 'proper' dynamical class."[68]

The isomorphism of the process of self-organization may be an important factor in creating a bridge between the micro and macro domain and resolving the problem of the fundamental division in contemporary physics. It suggests a vision of reality as a nested hierarchy of cascading levels of organization which, although fundamentally different and irreducible to each other, are nevertheless intimately related in their patterns of emergence and transformation.

Halzhey observes:

> While this [universal applicability of the theory of self-organization] would also apply, for instance, to the fundamental laws of physics, applicability on all scales engenders in the case of self-organization an even more sublime image, namely the image of an infinite series of hierarchically nested levels that are analogous to each other. . . . The homology across different scales pertains therefore rather to processes of emergence or to the transition between levels than to the levels themselves.[69]

These observations imply that the exploration of interdisciplinary paths of theorizing in the context of theory of self-organization and other theoretical perspectives associated with it may offer a key to resolving paradoxes in contemporary physics which result from the radical differentiation of the micro and the macro domain, including the paradox of Schrödinger's cat. T. Palmer, for example, points in this direction in his discussion of the patterns of self-similar upscale cascading and the so-called "meteorological Butterfly effect" which he sees as offering a possibility "of overcoming some of the objections to marrying chaos theory and quantum theory."[70]

This discussion of paradoxes in contemporary physics suggests that paradoxes are intimately associated with the process of construction. They appear when a newly emergent structure clashes with its constituent substructures. This clash is largely due to the fact that the newly emergent structure, as a combination of structures involved in its creation, is much more powerful, that is, has a much greater combinatorial capability, and hence is capable of constructing more statements, than the latter; hence paradoxes. Paradoxes that we encounter appear as a result of our advances in knowledge and are produced by the conflict between the emergent cognitive structure and those structures that are involved in its emergence. This conflict can be resolved only through the process of equilibration, or adaptation. The result of such equilibration will be further advances in knowledge, new sources of disequilibrium, and, inevitably, other paradoxes. As I have argued elsewhere,[71] conservation and regulation are the two processes that play a key role in this regard. Any knowledge structure requires regulation to conserve itself. And such regulation will eventually evolve into a new structure that will in turn also require regulation. The construction of knowledge is, indeed, an infinite process.

Point-Of-View Invariance

The conception that emphasizes the constructed, rather than reflective nature of knowledge does not necessarily have to lead one, as it did Post-Modernists, to a pessimistic conclusion that there can be no correspondence between our knowledge and reality and that objective knowledge is in principle unattainable. It is worth noting, that the dualism underlying the dichotomy between the object and the subject is largely a result of the currently prevalent approach to knowledge which pays little attention to the process of construction. Both subject and object are merely aspects of the process of construction that appear to be opposed to each other when the process of construction is disregarded. The question whether objective knowledge is possible is really a false question, indeed, generated by this dualistic perspective. Our experience confirms that it is possible to establish correspondences between what we know and what is out there. The real question is not whether we can have objective knowledge. The real question is: How do we attain such knowledge? What do we do to produce knowledge that corresponds to reality?

The most important criterion of objective knowledge is invariance, or point-of-view invariance (POVI). As Victor Stenger points out, "If the models of physics are to describe observations based on an objective reality, then those models cannot depend of the point of view of the observer. This suggests the principle of point-of-view invariance . . . ."[72] Newton introduced POVI with regard to the direction of motion. His laws are invariant in relation to the direction. One of the great achievements of Einstein was to generalize POVI to space-time. In his view, no space-time frame is preferable to any other, and the constancy of the speed of light was the guarantee of this invariability. However, while Einstein established the principle of invariability of space-time, which implicitly recognized the agency of the knower in constructing space-time frames, he did not fully embrace this agency. He did not recognize that the act of knowing affected all systems and that the condition of observing a system "without in any way affecting it" was an impossibility.

      QT reflects keen awareness of the impact that the knower has on the object of knowing. However, without a clear understanding of the processes involved in knowing, its epistemology could appear unacceptably subjective. The notion that reality could only be known subjectively was totally unacceptable to the creators of QT. They had to incorporate objectivity into their theory. However, the only conception of objectivity that was available to them was a classical one, that is, the one that did not recognize the connection between the agent and the object of knowing. The Copenhageners certainly could not apply this conception of objectivity to quantum objects since that would mean a rejection of their most important theoretical innovation. Therefore, they applied it to the conditions of the experiment set up by the agent. They required a classical description of the conditions of a quantum experiment. But what would constitute such description? The classical description of the object is a description that requires isolation of the object from the knower. By analogy, a classical description of the conditions of an experiment should also presuppose isolation of these conditions from the knower. But how such isolation is supposed to be achieved? How are we supposed to remove subjectivity from the knowledge of the conditions of the experiment? Could we remove our subjective

biases from our descriptions? Wouldn't such claims of removal simply place the subject with his or her biases into a transcendent position? And wouldn't further repetitions of this operation of removal simply lead to an infinite regress? These questions suggest that while QT recognized the agency of the knower, it did not successfully generalize POVI to all action. It did not establish a framework which would not privilege one action over another; it does not incorporate the knower into the process of knowing.

Nothing is more fundamental to reality than change, and change is the product of action. Therefore action is the most important property of reality. Following Hamilton's principle of stationary action, that is, action invariant in all cases, Basilio Catania, for example, sees kinship between physical action (for example, energy and momentum) and knowledge, or information. According to Catania,

> . . . action is the fundamental invariant describing any kind of change in the outer world of the observer, much the same as information is the fundamental invariant describing change in the inner world (mind) of the observer . . . . It therefore appears that action may be seen today as the unifying quantity of all physical quantities, much the same as Giovanni Giorgi saw energy playing that role in 1901.[73]

Catania, for example, finds that Plank's constant ℏ can be used to measure both actions—the physical change external to the knower and inner changes in knowledge. If, indeed, action is the most fundamental property of reality, then objective knowledge requires that POVI should be extended to action.

In order to extend POVI to action, one should find a point which is common to all actions and view reality from this point. The process of construction is fundamental to the evolution of reality. It is this process that generates change. As I have argued in one of my papers, the most important characteristic of this process is the equilibrium between equilibrium and disequilibrium. It is this equilibrium that allows the process of construction to move constantly forward without erupting into chaos. The locus of this equilibrium can be properly described as being at the "edge of chaos." I have also argued that it is precisely this balance that regulates the process of construction and therefore offers a position from which to reflect on the entire process. It allows the knower to reflect on one's own knowing, on the process of construction, and thus incorporate the knower into the process of knowing.[74] Therefore, in order to extend POVI to action one should view reality from the vantage point of this fundamental symmetry. This approach gives no preference to any frame of action or any knowledge-product of any action.

The focus on the process of construction helps to understand better why it is possible to establish correspondences between our knowledge and reality. First of all, in accordance with this theoretical perspective, the production of knowledge is isomorphic to other processes of construction that operate in reality. This view is totally consistent with the evolutionary approach to the study of reality. Cognitive processes emerge from biological ones which, in turn, developed from the processes that occur at the physical and chemical levels of organization of reality. What one author has described as "an unreasonable effectiveness of mathematics in the natural sciences" is a good example of this isomorphism.[75] Just as the processes of construction in nature are real, so are the constructions effected by our mind. Secondly, as Piaget has convincingly showed, our consciousness and our knowledge are products of our interaction with the real world and, therefore, are real.[76] These two important facts are what make correspondences between what we know and reality possible.

Possibility, however, is no guarantee. That is why there are a number of important criteria that have been established for attaining objective knowledge. Inclusiveness is one of them: the more inclusive is our interpretation of reality, the more it incorporates other theoretical perspectives as its specific cases, the greater the likelihood of its invariant objectivity. Criticality is another important criterion, that is, the more critical is our theory, the more it includes the knower into the process of interpreting reality, the more the knower is aware of his or her knowing biases, the more objective the interpretation is likely to be . Finally, there are well established practices of experimental and evidentiary confirmation.

The above considerations confirm a well-known empirical fact that we are capable of attaining knowledge that corresponds to the real world. Although the process of construction, and the production of knowledge is one of its manifestations, is infinite, its specific products are not. In view of this fact, it is possible to have finite and fully objective knowledge of finite objects.

Conclusion

As this paper has shown, paradoxes are essential products of the process of construction. Paradoxes announce the presence of a new and more powerful cognitive structure. It is the conflict between this new emergent structure and the structure which led to its emergence that is the source of the perturbation that we experience as paradox, that is, statements that are true and false at the same time—true if viewed from the point of view of the former but unjustifiable and hence perturbing when viewed from the point of view of the latter.

This view suggests that the only one way to deal with perturbations that create paradoxes is to move boldly forward and embrace the emergent structure. The structure that led to its emergence should be equilibrated with the emergent structure and become a particular case within its more powerful framework. Thus the resolution of the problem of paradox lies in a bold construction of new knowledge.

The theoretical vision outlined in this paper offers a new perspective on paradoxes and their role. Viewed from this perspective, paradoxes cease to appear as indications of structural limitations to human knowledge. Rather, they become powerful tools in the production of knowledge. This perspective allows the knower the freedom of critical insight, a capacity to inquire into the basis of one's own knowledge, and ultimately a better control over the process of construction and, hence, its greater efficiency. An understanding of the process of construction will allow us to become masters of our own creativity. By controlling this process we can learn to be creative when we want to, not only when we can. Rather than suffuse, as we often do, the production of knowledge in a futile and wasteful exercise of power, we can turn it into a more efficient, more cooperative, more orderly, and ultimately much more enjoyable process.

---

[1] See, for example, T. Williamson, Knowledge and its Limits, chap. 12, pp. 270-340.
[2] Roy Sorensen, A Brief History of the Paradox: Phylosophy and the Labyrinths of the Mind (Oxford: Oxford University Press, 2003), p. xii.
[3] R. M. Sainsbury, Paradoxes (Cambridge: Cambridge University Press, 1995), p. 1.
[4] Marianne W. Lewis, "Exploring Paradox: Toward a more Comprehensive Guide," Academy of Management Review, Vol. 25, No. 4 (2000), p. 760.
[5] According to most sources, the first formulation of this paradox is attributed to Eubalides of Miletus. See, for example, Sorensen, A Brief History of the Paradox, pp. 93-95.

[6] Roy A. Sorensen, Blindspots (Oxford: Clarendon Press, 1988).

[7] R. M. Sainsbury, Paradoxes (Cambridge: Cambridge University Press, 1995), p. 2 and p. 135.

[8] J. D. Ford and R. H. Backoff, "Organizational change in and out of dualities and paradox," in R. E. Guinn and K. S. Cameron, eds., Paradox and Transformation: Toward a Theory of Change in Organization and Management (Cambridge, MA: Ballinger, 1988), p. 89.

[9] Sainsbury, Paradoxes, p. 1.

[10] See Ernest Nagel and James R. Newman, Godel's Proof (New York: New York University Press, 1953). Godel himself did not spell out this argument which raises doubts, for example, about possibility of creating artificial intelligence—one of the rapidly developing areas in modern Computer Science. However, the fact that Nagel and Newman drew this conclusion shows that it is not unwarranted (Nagel and Newman, Gödel's Proof, pp. 100-01).

[11] I define cognitive structure as a set of interrelated cognitive operations—which can be any operation from reflexes to symbolic operations—used for construction of knowledge.

[12] Jean Piaget, The Origin of Intelligence in Children (Madison, Conn.: International Universities Press, 1998); Gennady Shkliarevsky, "The Paradox of Observing, Autopoiesis, and the Future of Social Sciences," Systems Research and Behavioral Science, vol. 24, issue 3 (May/June 2007), pp. 323-32.

[13] Online source at http://www.spaceandmotion.com/Physics-Richard-Feynman-QED.htm#Quotes.Richard.Feynman (accessed on October 20, 2008).

[14] John Cramer provides a detailed description of these interpretations in the addendum to his article "The Transactional Interpretation of Quantum Mechanics," Review of Modern Physics, vol. 58, No. 3 (July 1986), pp. 647-87.

[15] Niels Bohr, "Discussion with Einstein on Epistemological Problems of Atomic Physics," in P. A. Schilpp, Albert Einstein: Philospher-Scientist, 2. vols. (London: Cambridge University Press, 1982), vol. 1, p. 209 (emphasis in the original).

[16] N. P. Landsman, "When Champions Meet: Rethinking the Bohr-Einstein Debate," Studies in History and Philosophy of Modern Physics, vol. 37 (2006), p. 221, Henry P. Stapp, "Quantum Theory and the Role of Mind in Nature," arXiv:quant-ph/0103043 v1 9March 2001 (accessed October 3, 2008), p. 6.

[17] Gennady Shkliarevsky, "Deconstructing the Quantum Debate: Toward a Non-Classical Epistemology," arXiv:0809.1788v1 [physics.hist-ph] September 10 2008.

[18] Cramer, "Transactional Interpretation," p. 681.

[19] Cramer, for example, argues convincingly that quantum mechanics produces paradoxes when the principle of invariability is brought in. It is not invariable. There is a preferred frame. Laws must be the same in all frames (Cramer, "Transactional Interpretation," pp. 656-57). He shows incompatibility between General Relativity and quantum mechanics which does not observe relativistic invariance.

[20] As quoted in Cramer, "Transactional Interpretation," p. 657.

[21] R. P. Feynman, The Character of Physical Law (Cambridge, MA: MIT Press, 1967), p. 129.

[22] Adrian Kent, "Night thoughts of a quantum physicist," Philosophical Transactions of the Royal Society of London A, vol. 358 (2000), p. 77.

[23] Cramer, "Transactional Interpretation," p. 658.

[24] David Pratt, "Review of Stanley Jeffers, et al., eds., Fean-Pierre Vigier and the Stochastic Interpretation of Quantum Mechanics (Montreal: Apeiron, 2000)," Journal of Scientific Exploration, vol. 1, No. 2 (2002), p. 283.

[25] Harvey B. Brown and David Wallace, "Solving the Measurement Problem: deBroglie-Bohm Loses out to Everette," arXiv: quant-ph/0403094v1 12 March 2004, p. 8 (accessed Nov. 5, 2008).

[26] Pratt, "Review," p. 283.

[27] K. K. Thechendath, "Stephen Hawking and Schrödinger's Cat," Social Scientist, vol. 22, Nos., 7-8 (July-August 1994), p. 107.

[28] F. Laloë, "Do we really understand quantum mechanics? Strange correlations, paradoxes, and theories," American Journal of Physics, vol. 69 (6) (June 2001), .p. 660.

[29] As quoted in John Small, "Why do Quantum Systems Implement Self-Referential Logic? A Simple Question with a Catastrophic Answer," in D. M. Dubois, ed., Computing Anticipatory Systems: CASYS'05: Seventh International Conference (American Institute of Physics, 2006), p. 167.

[30] Kent, "Night thoughts," p. 77.

[31] John G. Cramer, "The transactional interpretation of quantum mechanics," Reviews of Modern Physics, vol. 58, No. 3 (July 1986), pp. 647-87.

[32] Cramer, "Transactional Interpretation," p. 665.

[33] Ibid., p. 663.

[34] Ibid., p. 662.

[35] Ibid., p. 663.

[36] Ibid., p. 645.

[37] Ibid., p. 668.

[38] Ibid., p. 648.

[39] Pavel V. Kurakin, George G. Malinetskii, Howard Brown, "Conversational (dialogue) model of quantum transitions," arXiv:quant-ph/0504088v2 14 April 2005 (accessed February 12, 2008); Pavel V. Kurakin and George G. Malinetskii, "A Simple Hypothesis on the Origin and Physical Nature of Quantum Superposition of States," arXiv:physics/0505120v1 [physics.gen-ph] 17 May 2005 (accessed February 12, 2008).

[40] Kurakin, et al., "Conversational model," p. 1.

[41] As quoted in John Small, "Why Do Quantum Systems Implement Self-Referential Logic? A Simple Quastion with a Catastrophic Answer," presented at the Seventh International Conference of Computing Anticipatory Systems CASYS '05 (American Institute of Physics, 2006), p. 167.

[42] Shkliarevsky, "Deconstructing the quantum debate," especially pp. 11-12.

[43] Piaget, The Origin of Intelligence in Children.

[44] Sengupta, "Quantum non-locality and complex holism," Nonlinear Analysis, vol. 69 (2008), p. 1009.

[45] Vladimir Manasson, "Are Particles Self-Organized Systems?" arXiv:0803.3300—1 [physics.gen-ph] 23 March, 2008 (accessed August 23, 2008).

[46] Xiao Gang Wen, "An Introduction to Quantum- Order, String-Net Condensation, and Emergence of Light and Fermions," http://dao.mir.edu/-wen (accessed July 9, 2008), p. 1.

[47] Ignazio Licata, "Emergence and Computation at the Edge of Classical and Quantum Systems." arXiv:0711.2973v1 [quant-ph] (accessed January 12, 2009).  See Vladimir Manasson, "Are Particles Self-Organized Systems?"  arXiv:0803.3300—1 [physics.gen-ph]  23 March, 2008 (accessed August 23, 2008); A. Sengupta, "Quantum non-locality and complex holism," Nonlinear Analysis, vol. 69 (2008), pp. 1000-1010, Maria Koleva, "Self-Organization and Finite Velocity of Transmitting Substance and Energy through Space-Time," arXiv:nlin/0601048v1 [nlin.AO] (accessed October 23, 2008); Christoph Maschler, et al., "Entaglement driven self-organization via a quantum seesaw mechanism," arXiv:quant-ph/0512101v1  13 Dec 2005 (accessed November 21, 2008); Matti Pitkanen, "Self-Organization and quantum jumps," http://64.233.169.132/search?q=cache:91BDngW9gtEJ:www.helsinki.fi/~matpitka/selfog (accessed December 18, 2008).

[48] Pavel V. Kurakin, George G. Malinetskii, "How Bees Can Possibly Explain Quantum Paradoxes," http://www.chronos.msu.ru/EREPORTS/kurakin_how.pdf   ( accessed December 1, 2008); also Pavel V. Kurakin, George G. Malinetskii, "A Simple Hypothesis on the Origin and Physical Nature of Quantum Superposition of States," arXiv:physics/0505120v1 [physics.gen-ph]  (accessed December 1, 2008).

[49] Lain D. Couzin and Jens Krause, "Self-organizatin and Collective Behavior of Vertebrates," http://www.princeton.edu/~icouzin/Couzin&Krause.pdf, p. 1 (accessed July 10, 2008).

[50] James Odell, "Agents and Beyond:  A Flock is not a Bird," http://jeffsutherland.com/oopsla98/odellagents.pdf, p. 6 (accessed on March 4, 2009).

[51] See, for example, Claus Emmeche, "Explaining Emergence:  Towards an Ontology of Levels," http://www.nbi.dk/~emmeche/coPubl/97e.EKS/emerg.html, p. 18 (accessed June 20, 2008).  Emmeche also considers that complex systems "are accessible for formal and scientific treatment (ibid., p. 22).

[52] See for example, M. Polany, "Life's Irreducible Structure," Science, No. 160, pp. 1308-13.

[53] Christoph Holzhey, "Self-organization at the Edge of Mysticism," a paper presented at Volkswagen Conference (Gainesville, Florida, 2004), http://www.mystik.uni-siegen.de/Texte/Holzhey2004a.pdf, p. 4 (accessed September 30, 2009).

[54] See, for example, Iain D. Couzin and Jens Krause, "Self-organization and collective behavior in vertebrates."

[55] Cleaire Detrain and Jean-Louis Deneubourg, "Self-organized structures in a superorganism:  do ants 'behave' like molecules," Physics of Life Reviews. No. 3 (2006), pp. 162-87; Mark Fleischer, "Foundations of Swarm Intelligence:  From Principles to Practice," arXiv:nlin/0502003v1 [nlin.AO]  2 Feb 2005 (accessed on January 4, 2009]; Simon Garnier, et al., "The biological principles of swarm intelligence," Swarm Intelligence, No. 1 (2007),  pp. 3-31.

[56] Erzilla Lozneanu, et al., "Ball Lightning as a Self-Organized Complexity," arXiv:0708.4064v1 [nlin.PS]  (accessed September 23, 2008).

[57] T. N. Palmer, "Quantum Reality, Complex Numbers and the Meteorological Butterfly Effect," arXiv:quant-ph/0404041v2 (accessed November 3, 2008).

[58] Steven V. Viscido, et al., "Factors influencing the structure and maintenance of fish schools," Ecological Modelling, vol. 206 (2007), pp. 153-165.

[59] Vito Pirelli, et al., "Non-locality all the way through: Emergent Global Constraints in the Italian Morphological Lexicon," http://www.aclweb.org/anthology/W/W04/W04-0102.pdf (accessed July 20, 2008).

[60] See my "The Paradox of Observing, Autopoiesis, and the Future of Social Sciences."

[61] Eric Bonabeau et al., "Self-organization in Social Insects," http://www.santafe.edu/research/publications/workingpapers/97-04-032.pdf, p. 5.

[62] Couzin and Krause, "Self-Organization," p. 14; D. Radakov, *Schooling and Ecology of Fish* (New York: Wiley, 1973).

[63] See, for example, Couzin and Krause, "Self-organization and collective behavior in vertebrates."

[64] See, for example, Simon Garnier, et al., "The biological principles of swarm intelligence," Swarm Intelligence, vol. 1, no. 1 (June 2007), pp. 3-31.

[65] Vito Pirelli, et al., "Non-locality all the way through: Emergent Global Constraints in the Italian Morphological Lexicon."

[66] Sengupta, "Quantum Non-Locality," p. 1004.

[67] Brian D. Josephson, "Beyond Quantum Theory: A Realist Psycho-Biological Interpretation of Reality Revisited," BioSystems, vol. 64 (2002), p. 44

[68] Manasson, "Are Particles Self-Organized Systems," p. 2.

[69] Halzhey, "Self-organization," p. 1 and 2.

[70] T. N. Palmer, "Quantum Reality, Complex Numbers and the Meteorological Butterfly Effect," p. 17.

[71] Shkliarevsky, "The Paradox of Observing."

[72] Victor Stenger, "Where Do the Laws of Physics Come From?" http://www.colorado.edu/philosophy/vstenger/Nothing/Laws.pdf, p. 1 (accessed June 20, 2009).

[73] Basilio Catania, "The Action Unit as a Primary Unit in SI," Lecture at the International Meeting "Giovanni Giorgi and his Contribution to Electrical Metrology," Polytechnic of Turin, 21-22 September 1988, http://www.chezbasilio.it/immagini/Action_unit.pdf, p. 3 (accessed May 19, 2008).

[74] G. Shkliarevsky, "The Paradox of Observing, Autopoiesis, and the Future of Social Sciences."

[75] Karl Svozil, "Physical Unknowables," contribution to the International symposium "Horizons of Truth" (2006), arXiv:physics/0701163v2 [physics.gen-ph], p. 2 (accessed June 12, 2008).

[76] Piaget, The Origin of Intelligence in Children.